\documentstyle[aps,prl]{revtex}

\begin{document}
\draft
\title{Isospin effects on squeeze-out flow in heavy-ion collisions}
\author{Feng-Shou Zhang$^{1,2,5}$, Lie-Wen Chen$^{1,2,3}$, Zhi-Yuan Zhu$^{1,4}$}
\address{$^1$ Center of Theoretical Nuclear Physics, National Laboratory of Heavy Ion%
\\
Accelerator, Lanzhou 730000, China\\
$^2$ Institute of Modern Physics, Academia Sinica, P.O. Box 31, Lanzhou
730000, China\\
$^3$ Department of Applied Physics, Shanghai Jiao Tong University, Shanghai
200030, China\\
$^4$ Shanghai Institute of Nuclear Research, Academia Sinica, Shanghai
201800, China\\
$^5$ CCAST (World Laboratory), P.O. Box 8730, Beijing 100080, China}
\maketitle

\begin{abstract}
The squeeze-out flow in reactions of $^{124}$Sn + $^{124}$Sn and $^{124}$Ba
+ $^{124}$Ba at different incident energies for different impact parameters
is investigated by means of an isospin-dependent quantum molecular dynamics
model. For the first time, it is found that the more neutron-rich system ($%
^{124}$Sn + $^{124}$Sn) exhibits weaker squeeze-out flow. This isospin
dependence of the squeeze-out flow is shown to mainly result from the
isospin dependence of nucleon-nucleon cross section and the symmetry energy.
\end{abstract}


The recent advance in radioactive nuclear beam (RNB) physics provides people
a unique opportunity to investigate isospin effects in heavy-ion collisions
(HIC's)\cite{LbaIJMP98,DitoroPNPP99}. Isospin effects on directed flow\cite
{LbaPRL96,PakPRL971,ClwPRC98}, radial flow\cite{ClwPLB99} and rotational flow%
\cite{ClwPRC00} in HIC's at intermediate energies have been explored
theoretically and/or experimentally, and it is indicated that the isospin
dependence of nuclear collective flow may provide people some important
information about the isospin-dependent nuclear EOS, particularly the
symmetry energy, and the isospin-dependent in-medium {\sl N-N} cross
section. The squeeze-out flow\cite{GutbrodPLB89} is of special interest
since it comes from the only direction where the nuclear matter might escape
without being hindered by the presence of the cold spectator remnants and
thus a less disturbed information on the matter of high density and
temperature is expected. Therefore, it is very significant to explore
isospin effects on the squeeze-out flow.

In this Short note we report results of the first theoretical study on the
squeeze-out flow from reactions of $^{124}$Sn + $^{124}$Sn and $^{124}$Ba + $%
^{124}$Ba at different energies for different impact parameters within the
framework of an isospin-dependent quantum molecular dynamics (IDQMD) model
which includes the symmetry energy, Coulomb interaction, isospin-dependent
experimental $N$-$N$ cross sections, and particularly the isospin-dependent
Pauli blocking\cite{ClwPRC98,ClwJPG97}. In the initialization process of the
IDQMD model, the neutron and proton are distinguished from each other and
meanwhile the nonphysical rotations in the initialized nuclei have been
removed\cite{ClwPRC00}. In the IDQMD model, the nuclear mean field can be
parameterized by 
\begin{eqnarray}
U({\rho },{\tau }_z) &=&{\alpha }({\rho }/{\rho }_0)+{\beta }({\rho }/{\rho }%
_0)^\gamma +{\frac 12}(1-{\tau }_z)V_c  \nonumber \\
&&\ \ \ +C\frac{{\rho }_n-{\rho }_p}{{\rho }_0}{\tau }_z+U^{Yuk},
\label{EqMF}
\end{eqnarray}
with ${\rho }_0$ the normal nuclear matter density (here is 0.16 fm$^{-3}$); 
${\rho }$, ${\rho }_n$, and ${\rho }_p$ are the total, neutron, and proton
interaction densities, respectively; ${\tau }_z$ is the $z$th component of
the isospin degree of freedom, which equals $1$ or $-1$ for neutrons or
protons, respectively; $C$ is the symmetry strength; $V_c$ is the Coulomb
potential; and $U^{Yuk}$ is the finite range Yukawa (surface) potential
which will vanish for infinite nuclear matter. The forms and parameters of
Eq. (\ref{EqMF}) can be found in Ref.\cite{ClwPLB99}. The IDQMD is different
from the so-called IQMD (Isospin-QMD) model\cite{Ha89,Bass95,Ha98} by the
Pauli blocking, the initialization process, and construction of fragment.
This model has been used recently to explain successfully several phenomena
in HIC's at intermediate energies, which depend on the isospin of the
reaction system\cite{ClwPRC98,ClwPLB99,ClwPRC00,ClwJPG97,ZfsPRC99}. In the
present calculations, the so-called soft EOS with an incompressibility of $%
K=200$ MeV is used and the symmetry strength $C=32$ MeV without particular
consideration.

In the QMD model\cite{AichelinPR91}, the reaction plane is known $a\ priori$
and it is defined as the $x$-$z$ plane ($z$-axis corresponds to the beam
direction). The azimuthal angle with respect to the reaction plane can be
written as

\begin{equation}
\phi =\arctan (P_y/P_x),  \label{EqPhi}
\end{equation}
where $P_x$ and $P_y$ are the $x$- and $y$-components of nucleon momentum in
the center-of-mass (c.m.) system. Many studies have shown that the azimuthal
distribution in HIC's can be fitted well by the Legendre polynomial up to
the second order\cite{ReisdorfARNPS97}, i.e.,

\begin{equation}
dN/d\phi =c(1+a_1\cos (\phi )+a_2\cos (2\phi )).  \label{EqLeg}
\end{equation}
The coefficient $a_1$ represents the strength of the in-plane directed flow
while a positive $a_2$ reflects the strength of the rotational collective
motion (the azimuthal distribution peaks at $\phi =0^{\circ }$ and $\pm
180^{\circ }$ simultaneously) and negative one (the azimuthal distribution
peaks at $\phi =$ $\pm 90^{\circ }$ ) the out-of-plane squeeze-out. For the
mid-rapidity azimuthal distribution, the ratio $R_N$,

\begin{equation}
R_N=\frac{dN/d\phi (90^{\circ })+dN/d\phi (-90^{\circ })}{dN/d\phi (0^{\circ
})+dN/d\phi (180^{\circ })}=\frac{1-a_2}{1+a_2},  \label{EqSq}
\end{equation}
measures the strength of the squeeze-out flow in a quantitative way. A value 
$R_N>1$ corresponds to a preferential out-of-plane emission. It has been
shown that $R_N$ depends on the transverse momenta of nucleons\cite
{HartnackPLB94} and in the present study the normalized nucleon transverse
momentum $P_t/P_{proj}$ (where $P_t$ and $P_{proj}$ represent the transverse
momenta of nucleon and the projectile momentum in the c.m. system,
respectively) is limited to $P_t/P_{proj}\geq 0.5$ for simplicity.
Meanwhile, the mid-rapidity is defined as a narrow region around the c.m.
rapidity by applying the condition $-0.25\leq (y/y_{\text{proj}})_{\text{c.m.%
}}\leq 0.25$, where $(y/y_{\text{proj}})_{\text{c.m.}}$ is the reduced c.m.
rapidity. In addition, it has been shown that $R_N$ depends on the fragment
mass\cite{HartnackPLB94} and in this work the calculated results are
obtained from all nucleons entering the analysis to accumulate the numerical
statistics.

Fig. 1 displays the IDQMD model prediction of $R_N$ as a function of
incident energy for reactions of $^{124}$Sn + $^{124}$Sn and $^{124}$Ba + $%
^{124}$Ba at impact parameter $b$=4 fm. The errors shown are the statistical
errors resulting from the Legendre polynomial fits. It is indicated in Fig.
1 that the $R_N$ value increases with increment of the incident energy and
then saturate or decrease at higher incident energies, which is in agreement
with the recent experimental results\cite{BastidNPA97,CrochetNPA97}. The
reduction of the $R_N$ value at higher incident energies is easy to
understand by the shadowing effects of the cold nuclear matter surrounding
the participant zone which decrease because the projectile spectators escape
faster. A more interesting feature in Fig. 1 is that the more neutron-rich
system $^{124}$Sn + $^{124}$Sn displays systematically smaller $R_N$ values
than the system $^{124}$Ba + $^{124}$Ba, which implies that there exists a
strong isospin dependence of the squeeze-out flow, namely, the more
neutron-rich system displays weaker squeeze-out flow. Meanwhile, one can
find that the isospin dependence of the squeeze-out flow decreases as the
incident energy increases, which may be due to the fact that the isospin
dependence of the {\sl N-N} cross section disappears at higher incident
energies.

In order to investigate further the isospin dependence of the squeeze-out
flow, we show in Fig. 2 the ratio $R_N$ as a function of impact parameter
for reactions of $^{124}$Sn + $^{124}$Sn and $^{124}$Ba + $^{124}$Ba at 350
MeV/nucleon. The errors shown are the statistical errors resulting from the
Legendre polynomial fits. Similarly, a strong isospin dependence of the
squeeze-out flow is observed once again, namely, the more neutron-rich
system $^{124}$Sn + $^{124}$Sn displays systematically smaller $R_N$ values
than the system $^{124}$Ba + $^{124}$Ba at different impact parameters. In
particular, this isospin dependence is more pronounced in semi-central
collisions. In addition, it is indicated that the largest $R_N$ value is
obtained at $b$=6 fm, i.e., in semi-central collisions, which is in good
agreement with the recent experiment where a maximum of $R_N$ located at
about $b$=6 fm has been evidenced in reaction of Au+Au at 400 MeV/nucleon%
\cite{BastidNPA97,CrochetNPA97}. This phenomenon can be explained by an
expansion shadowing scenario\cite{AichelinPR91} where the expansion of the
participant matter is rescattered by the cold target or projectile spectator.

It is important to investigate the influence of the symmetry energy and
isospin-dependent {\sl N-N} cross section on the squeeze-out flow since the
isospin dependence of the squeeze-out flow may result from the competition
among several mechanisms in the isospin-dependent reaction dynamics, such as
the symmetry energy, isospin-dependent {\sl N-N} cross sections, and so on.

Using different symmetry energy strength $C$ and parametrizations of {\sl N-N%
} cross sections, we show in Fig. 3 the IDQMD model predicted normalized
azimuthal distribution from $^{124}$Sn + $^{124}$Sn (solid circles) and $%
^{124}$Ba + $^{124}$Ba (open circles) at 350 MeV/nucleon and $b=6$ fm for
mid-rapidity nucleons. Meanwhile, the results of Legendre polynomial fits
according to Eq. (\ref{EqLeg}) for $^{124}$Sn + $^{124}$Sn (solid line) and $%
^{124}$Ba + $^{124}$Ba (dashed line) as well as the resulting $a_1$ and $a_2$
are also included in Fig. 3. For the results shown in Fig. 3 (a) we use $%
C=32 $ MeV and experimental {\sl N-N} cross section $\sigma _{\text{exp}}$
which is isospin dependent. In Fig. 3 (b) we use $C=0$ (no symmetry energy)
and $\sigma _{\text{exp}}$. The case of using $C=32$ MeV and Cugnon's {\sl %
N-N} cross section $\sigma _{\text{Cug}}$ which is isospin independent, is
plotted in Fig. 3 (c).

One can see from Fig. 3 that the azimuthal distribution exists minima at $%
\phi =0^{\circ }$ and $\pm 180^{\circ }$ (i.e., in the in-plane $P_x$%
-direction) and maxima at $\phi =$ $\pm 90^{\circ }$ (i.e., in out-of-plane $%
P_y$-direction which is perpendicular to the reaction plane). These features
imply that at mid-rapidity more nucleons are squeezed out perpendicular to
the reaction plane than in the reaction plane. In order to see more clearly
the isospin effects on the squeeze-out flow, we give in Table 1 the $R_N$
values extracted from $a_2$ in Fig. 3 for different cases. The errors shown
are the statistical errors resulting from the Legendre polynomial fits. One
can see from Table 1 that both the symmetry energy and the isospin-dependent 
{\sl N-N} cross section enhance the strength of the squeeze-out flow but the
latter enhances it more strongly. Particularly, it is indicated that the
influence of $\sigma _{\text{exp}}$ on system $^{124}$Ba + $^{124}$Ba is
stronger than that on the system $^{124}$Sn + $^{124}$Sn, which is easy to
understand since the neutron-proton cross section is about three times
larger than the neutron-neutron or proton-proton cross section for $\sigma _{%
\text{exp}}$ at energy of about 350 MeV/nucleon, which results in more {\sl %
N-N} collisions for $^{124}$Ba + $^{124}$Ba. On the other hand, the symmetry
energy is generally repulsive and the change of symmetry strength $C$ might
modify the equation of state and thus the squeeze-out flow. From above
analysis, one can conclude that the isospin dependence of the squeeze-out
flow seems to mainly result from the isospin dependence of {\sl N-N} cross
section and the symmetry potential has less influence on it. In above
calculations, only the free-space {\sl N-N} cross sections are adopted and
the in-medium effect is only simulated by the Pauli blocking. However, the
in-medium {\sl N-N} cross sections and their isospin dependence might be
strongly density dependent\cite{LgqPRC94,AlmNPA95}. The isospin dependence
of the squeeze-out flow may provide a unique opportunity to study the
isospin dependent in-medium {\sl N-N} cross sections.

In summary, by using the IDQMD model, we studied for the first time the
out-of-plane squeeze-out flow in reactions of $^{124}$Sn + $^{124}$Sn and $%
^{124}$Ba + $^{124}$Ba. A strong isospin dependence of squeeze-out flow has
been found, namely, the more neutron-rich system exhibits weaker squeeze-out
flow, which is shown to mainly result from the isospin dependence of {\sl N-N%
} cross section and the symmetry energy. Meanwhile, it is indicated that the
squeeze-out flow depends strongly on the impact parameter and incident
energy. Our study proposes that one can investigate the isospin-dependent
reaction dynamics by studying the isospin effects on the azimuthal
distribution and suggests that the isospin dependence of the squeeze-out
flow could be as a probe of the isospin-dependent in-medium {\sl N-N} cross
section.

This work was supported by the National Natural Science Foundation of China
under Grant NOs. 19875068 and 19847002, the Major State Basic Research
Development Program under Contract NO. G2000077407, and the Foundation of
the Chinese Academy of Sciences.

\section*{Figure captions}

\begin{description}
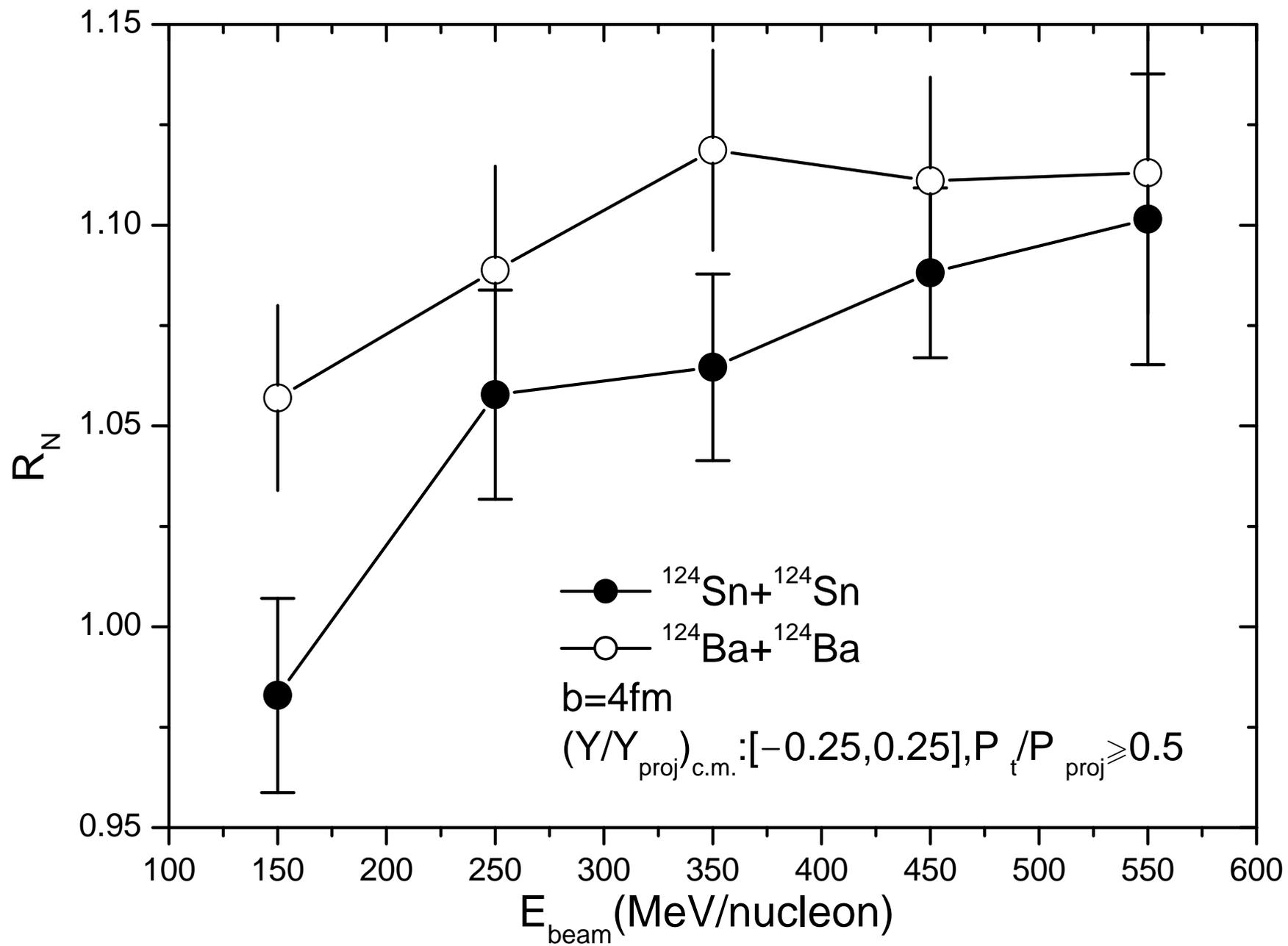
\item[Fig. 1]  The IDQMD model predicted $R_N$ as a function of incident
energy for reactions of $^{124}$Sn + $^{124}$Sn and $^{124}$Ba + $^{124}$Ba
at impact parameter $b$=4 fm. The lines are plotted to guide the eye.

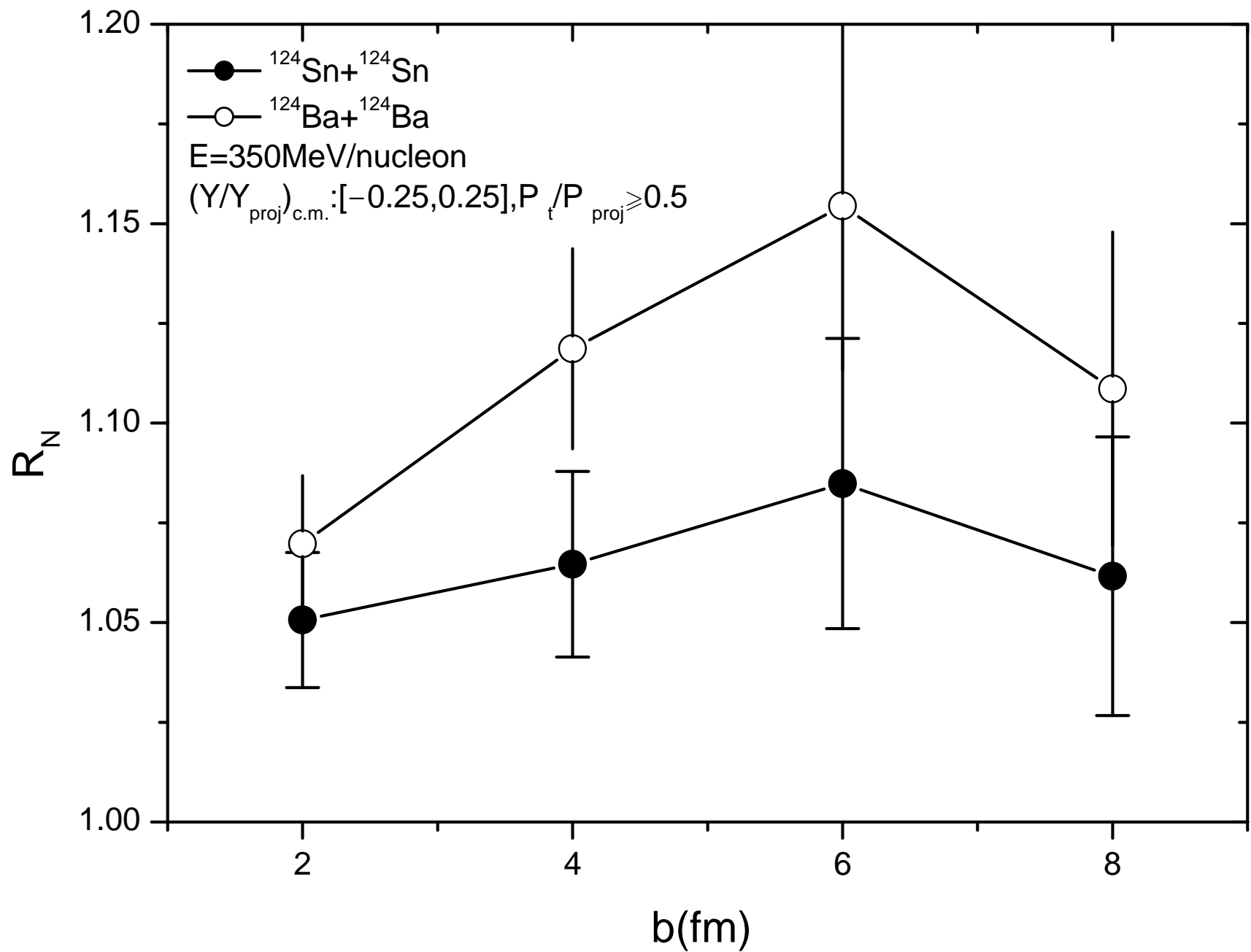
\item[Fig. 2]  The IDQMD model predicted $R_N$ as a function of impact
parameter for reactions of $^{124}$Sn + $^{124}$Sn and $^{124}$Ba + $^{124}$%
Ba at 350 MeV/nucleon. The lines are plotted to guide the eye.

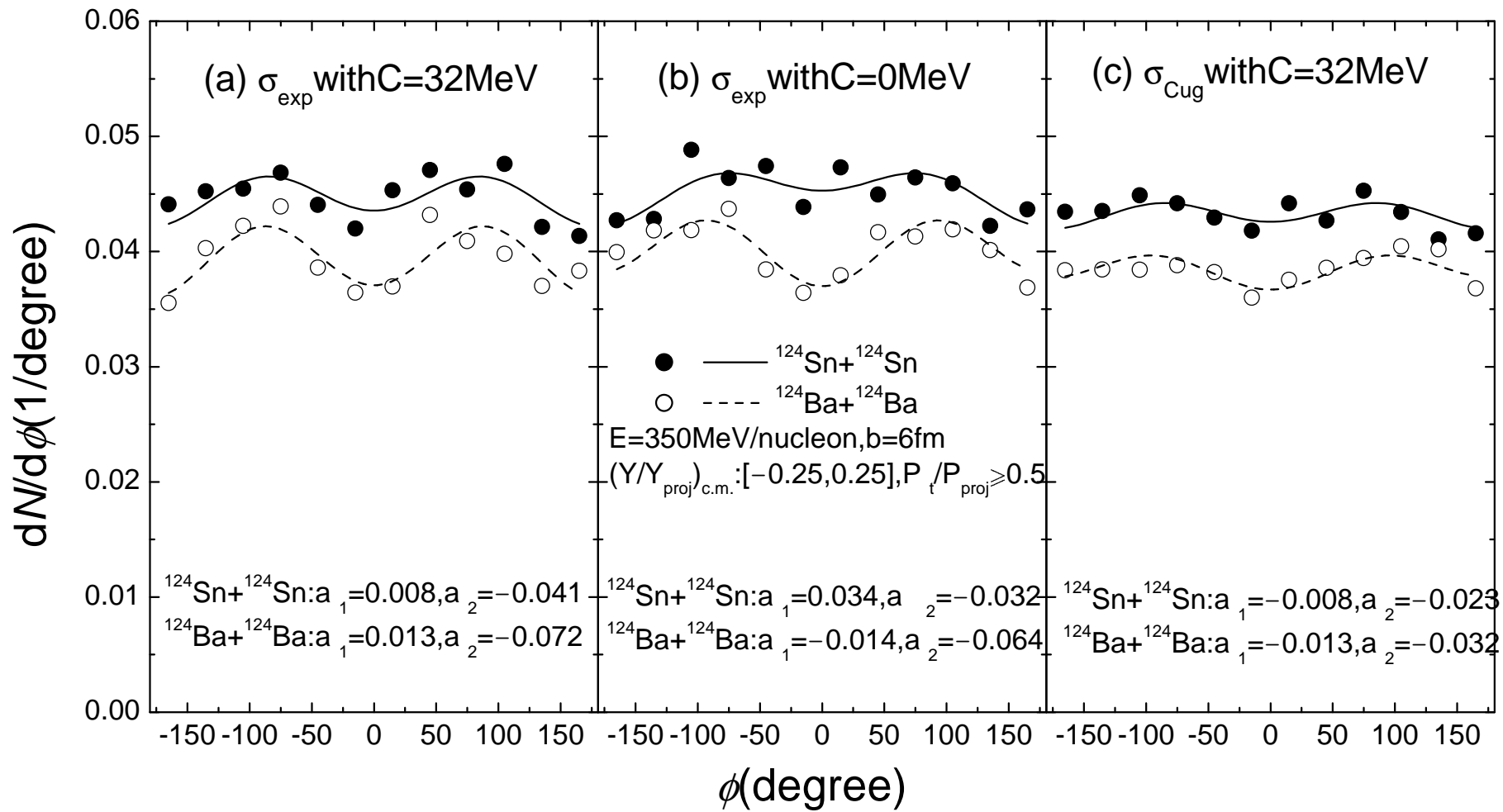
\item[Fig. 3]  The IDQMD model predicted normalized azimuthal distribution
for $^{124}$Sn + $^{124}$Sn (solid circles) and $^{124}$Ba + $^{124}$Ba
(open circles) at 350 MeV/nucleon and $b=6$ fm for mid-rapidity nucleons by
using different symmetry energy strength $C$ and parametrizations of {\sl N-N%
} cross sections: $C=32$ MeV with experimental {\sl N-N} cross section $%
\sigma _{\text{exp}}$ (a), $C=0$ (no symmetry energy) with $\sigma _{\text{%
exp}}$ (b), and $C=32$ MeV with Cugnon's {\sl N-N} cross section $\sigma _{%
\text{Cug}}$ (c). Meanwhile, the results of Legendre polynomial fits
according to Eq. (\ref{EqLeg}) for $^{124}$Sn + $^{124}$Sn (solid line) and $%
^{124}$Ba + $^{124}$Ba (dashed line) as well as the resulting $a_1$ and $a_2$
are also included.{\small \ }
\end{description}

\section*{Table captions}

\begin{description}
\item  Table 1: The $R_N$\ values at different situations (see text) for $%
^{124}$Sn + $^{124}$Sn and $^{124}$Ba + $^{124}$Ba at 350 MeV/nucleon and $%
b=6$\ fm.
\end{description}

\begin{center}
{\small 
\begin{tabular}{cccc}
\hline\hline
Reaction systems & $C$=32 MeV with ${\sigma }_{\text{exp}}$ & $C$=0 with ${%
\sigma }_{\text{exp}}$ & $C$=32 MeV with ${\sigma }_{\text{Cug}}$ \\ \hline
$^{124}$Sn + $^{124}$Sn & $1.086{\pm 0.036}$ & $1.066{\pm }0.029$ & $1.047{%
\pm }0.021$ \\ 
$^{124}$Ba + $^{124}$Ba & $1.155{\pm }0.041$ & $1.137{\pm }0.034$ & $1.066{%
\pm }0.023$ \\ \hline\hline
\end{tabular}
}
\end{center}

\end{document}